# Environmentally Stable Room Temperature Continuous Wave Lasing in Defect-passivated Perovskite

Jiyoung Moon[1], Masoud Alahbakhshi[2], Abouzar Gharajeh[2], Sunah Kwon[1], Ross Haroldson[3], Zhitong Li[2], Roberta Hawkins[1], Moon J Kim[1], Walter Hu[2], Anvar Zakhidov[3, 4], Qing Gu[2*]

[1]Department of Materials Science and Engineering, [2]Department of Electrical and Computer Engineering, and [3]Department of Physics, The University of Texas at Dallas, Richardson, TX 75080, [4]Department of Nanophotonics and Metamaterials, ITMO University, St. Petersburg, Moscow, Russia

*Corresponding author. Email: qing.gu@utdallas.edu

Metal halide perovskites have emerged as promising gain materials for on-chip lasers in photonic integrated circuits (PICs). However, stable continuous wave (CW) lasing behavior under optical pumping at room temperature – a prerequisite for electrically pumped lasing – has not yet been demonstrated. To achieve stable CW operation, we introduce a multiplex of strategies that include morphological, structural and interfacial engineering of $CH_3NH_3PbBr_3$ ($MAPbBr_3$) thin films to improve perovskite's intrinsic stability, as well as high quality cavity design to reduce the operational power. We demonstrate for the first time, over 90-minute-long green CW lasing with 9.4W/cm$^2$ threshold from a polycarbonate (PC)-defect-passivated, directly patterned $MAPbBr_3$ two-dimensional photonic crystal (PhC) cavity without any substrate cooling. We also show our approach's effectiveness on the performance of $MAPbBr_3$ under electrical excitation: we observe a seven-fold current efficiency enhancement by applying our strategies to a $MAPbBr_3$ LED. This work paves the way to the realization of electrically pumped lasing in perovskites.

An economical on-chip light source is a crucial component in high performance photonic integrated circuits (PICs) [1,2]. Although inorganic III-V and III-Nitride semiconductor lasers can be highly efficient and stable, they are typically off-chip due to the difficulty of their integration with the Silicon (Si) platform. Consequently, the system suffers from large coupling loss between the off-chip light source and Si chip at a high packaging expense. In the quest for alternative gain media for cost-effective and Si-compatible on-chip lasers, solution-processed and widely tunable hybrid perovskites have been proposed as the ideal candidate since the first demonstration of the perovskite laser in 2014 [3–5]. Moreover, high quantum efficiency, balanced ambipolar charge transfer, strong light absorption, and long carrier lifetime [6–12] prime perovskites for optoelectronic devices beyond lasers [13]. However, an electrically pumped perovskite laser has not yet been demonstrated. To achieve this ultimate goal, optically pumped lasing under continuous wave (CW) excitation without any substrate cooling is a crucial intermediate step. Lasing from various types of perovskite cavities such as the nanowire, microdisk, and photonic crystal (PhC) has been reported, but most were measured under pulsed laser excitation [14–23]. In the meantime, various room temperature perovskite lasers have also been reported, but they were all characterized under ultrafast pulsed excitation. Thus far, only a few studies investigated lasing in perovskites under CW excitation [24–27], and perovskite's poor thermal stability is believed to be the main culprit contributing to the lack of CW room temperature perovskite lasers [4,5]. To achieve this challenging operation condition, it is therefore critical to reduce device self-heating.

To reduce heat generation, the lasing threshold needs to be reduced, and one approach is to pattern the perovskite into a high-Q cavity. However, performing conventional lithography directly on perovskites is challenging due to their sensitivity to polar solvents, high temperatures and high-electron energies. Thus, new approaches to directly pattern perovskite thin films are in dire need [28]. By performing the solvent-free nanoimprint lithography (NIL) directly on $CH_3NH_3PbI_3$ ($MAPbI_3$), we previously reported CW room temperature lasing with 13W/cm$^2$ threshold from a $MAPbI_3$ distributed feedback laser [25]. Despite the very low threshold, the lasing action was only sustained for 240s around threshold, due to $MAPbI_3$'s fast degradation rate by exposure to heat, oxygen, moisture, and high photon energy generated by the UV pump laser [29–33]. For perovskite laser's insertion into PICs, much longer device lifetime is demanded. This can be achieved through laser cavity designs that promise even lower lasing threshold, inherent material stability improvement through structural or compositional changes to perovskite, and encapsulation to isolate the perovskite from the ambient atmosphere. In this work, we combine all of the above approaches and show for the first time, an environmentally stable perovskite laser that operates under CW pumping at room temperature, without any substrate cooling. This demonstration highlights (i) the synergistic effect of highly crystalline material, high Q cavity design, and defect passivation (ii) a facile, high-yielding, and controllable fabrication method that not only can be applied to other types of perovskite but also closely aligns with what industry seeks for mass production (iii) ultra-low lasing threshold and long lasing stability. We also show that our approaches improve device performance in a perovskite diode structure, making the prospect of an electrically pumped perovskite laser diode promising.

Results

The first step towards a high-quality laser cavity composed of perovskite thin film gain medium is the development of a well-controlled direct patterning method for perovskite. Our approach is to use the solvent-free NIL, the fundamental concept of which is depicted in Figure 1a. We perform NIL on $MAPbBr_3$ thin film to form a 2D PhC cavity (Figure 1a) and subsequently defect-passivate the $MAPbBr_3$ layer with polycarbonate (PC) (Figure 1b). Figure 1a and Figure 1b also show the scanning electron microscope (SEM) images of the $MAPbBr_3$ laser before and after PC passivation, respectively. Details of the fabrication process are described in Supplementary Part A. The $MAPbBr_3$ laser is characterized by optical excitation with a 355nm CW laser at room temperature with a humidity level of ~50% (see Supplementary Fig. 9 for the measurement apparatus).

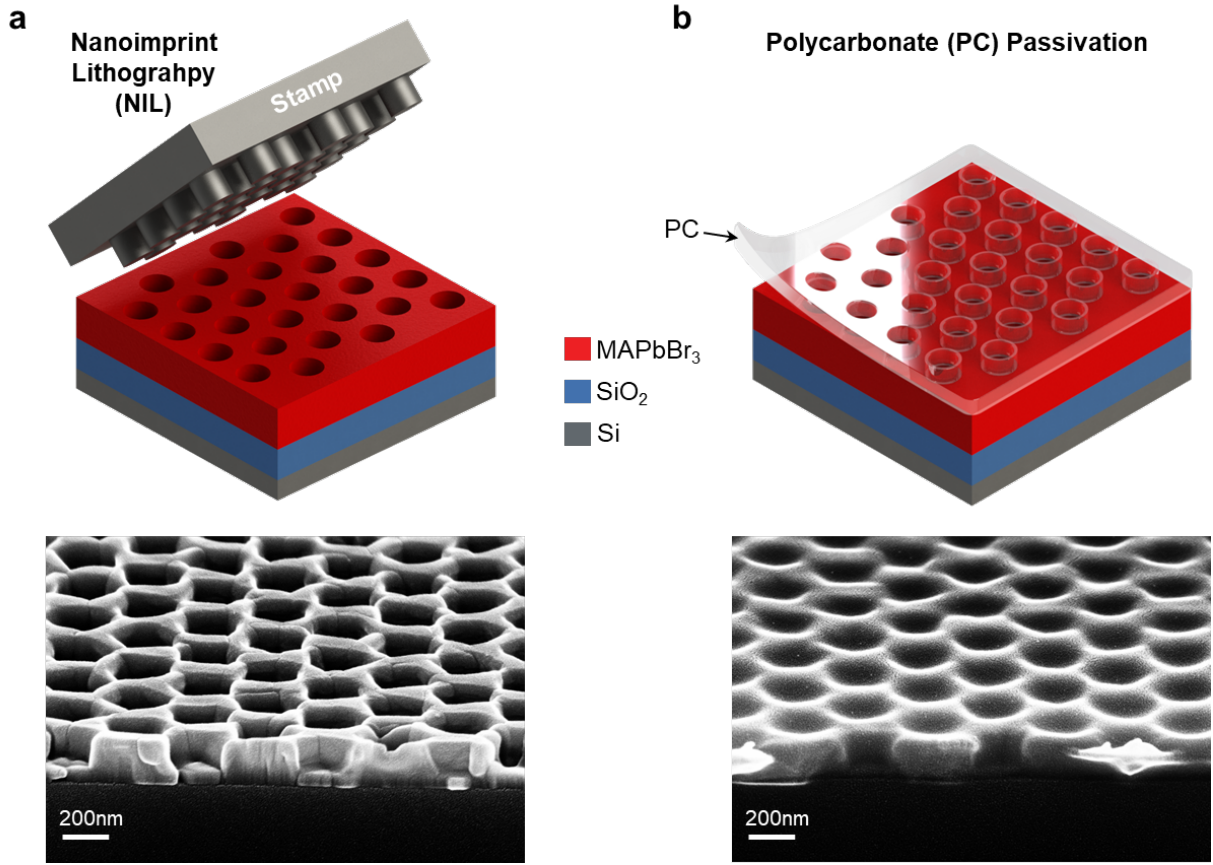

Figure 1 Stable MAPbBr$_3$ 2D PhC laser enabled by NIL and PC defect passivation. **a.** Schematic of the nanoimprint process on perovskite, and SEM image of the nanoimprinted MAPbBr$_3$ 2D PhC laser before PC defect passivation. **b.** Schematic of the defect passivation process, and SEM image of the MAPbBr$_3$ 2D PhC laser after PC defect passivation.

Figure 2a presents the light-in vs. light-out characteristics of the MAPbBr$_3$ 2D PhC laser, with a clear lasing threshold of 9.4W/cm$^2$. The lasing line at 553.9nm has a full width half maximum (FWHM) of 0.9nm at 12.9W/cm$^2$ (inset of Figure 2a, measured at 1.37P$_{th}$); however, linewidths as narrow as 0.6nm was measured in other samples of similar dimension. We only observe a lasing wavelength blue shift of <0.2nm over the entire pump power range, indicating insignificant device self-heating, thanks to the ultra-low lasing threshold that is a few orders of magnitude lower than most demonstrations in the literature. Figure 2b shows the evolution of the output power as a function of the pump power density. To tune the emission wavelength, in addition to varying the PhC dimension with the NIL stamp, we can also vary the perovskite residue layer thickness after NIL by varying the thin film thickness (See Supplementary Part B for optical simulations of the laser cavity). Figure 2c shows the controlled tuning of the lasing wavelength with PhC period 315~365nm, diameter 200~280nm, and perovskite residue thickness 0~150nm. To test lasing stability, we operate our laser at a CW pump power density of 11.14 W/cm$^2$ (1.18P$^{th}$) and record the lasing peak intensity until it drops to half of the initial value. As shown in Figure 1d, lasing is sustained without any surface treatment for ~60mins. Once the perovskite surface defects are passivated by a 50nm thick layer of PC film, lasing stability improves to over 90mins.

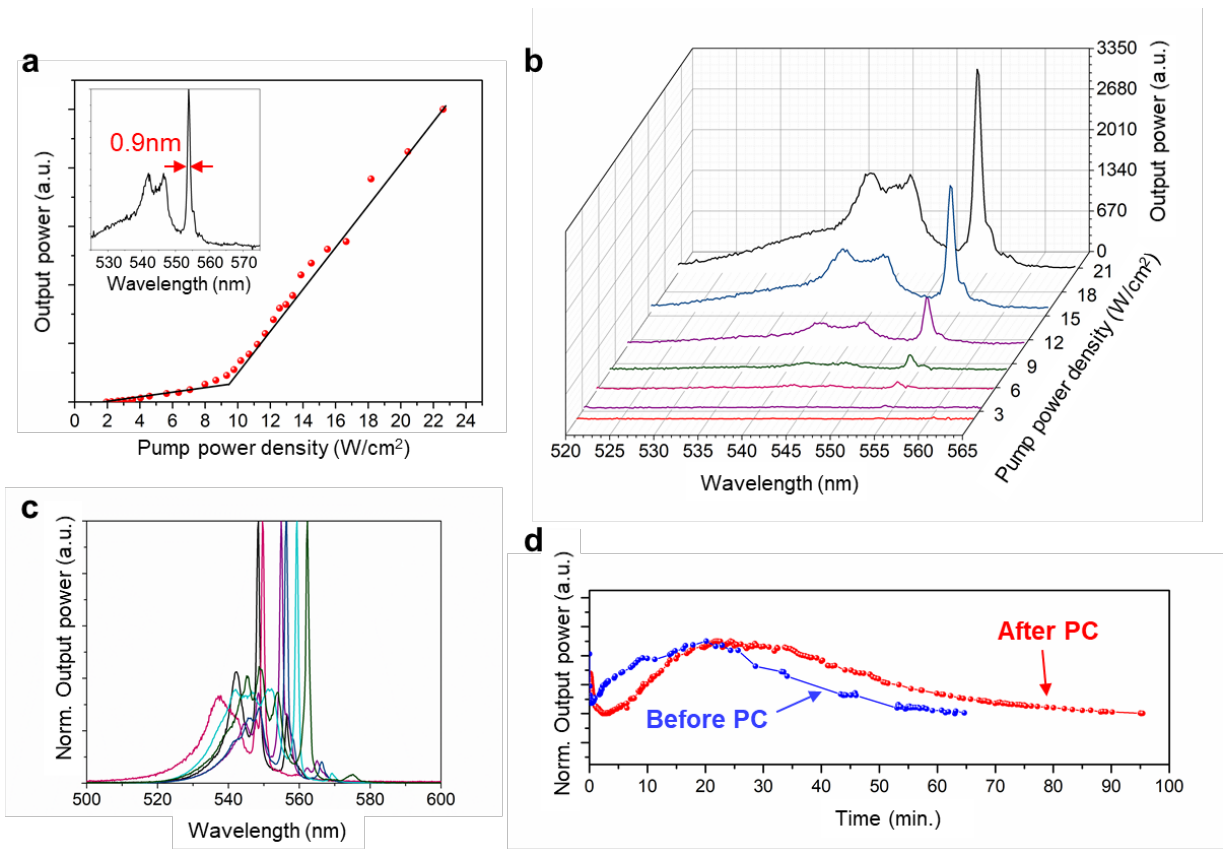

Figure 2 CW room temperature lasing characteristics of MAPbBr$_3$ 2D PhC lasers. **a.** Input-output characteristic of the laser (inset: zoom-in spectrum showing 0.9nm FWHM at 12.9W/cm$^2$). **b.** Output power as a function of pump power density. **c.** Lasing wavelength tuning via PhC dimension and perovskite residue thickness. **d.** Lasing stability before and after PC passivation.

Because perovskite thin film quality and device performance are intrinsically intertwined, improvements on perovskite's morphological properties such as enlargement of grain size [10,34], passivation of defects [9,35], and increase of the preferred crystal orientation [10] all lead to better device performance. We show that the imprinted MAPbBr$_3$ has excellent morphology and crystallinity, and PC passivation on the imprinted film leads to enhanced stability because PC not only blocks environmental hazards but also improves MAPbBr$_3$ film's surface quality by passivation of defects.

After NIL, the grain size of MAPbBr$_3$ is enlarged and the number of pinholes on the surface is reduced. To study these effects, we prepare two films: as-spin-coated MAPbBr$_3$ and nanoimprinted MAPbBr$_3$ with a flat stamp (i.e. flat NIL). First, we carry out scanning transmission electron microscopy annular bright field (STEM ABF) analysis. Comparing to SEM that detects secondary electrons only from sample's surface, the atomic-resolution STEM detects transmitted electrons from the bulk of the sample and shows thickness-dependent contrast from ABF images. STEM is therefore more effective in visually examining defects inside and on the surface of grains and grain boundaries. The STEM images in Figure 3a-b show that, after NIL, the number of bright regions that indicate crystallographically controlled voids decreases by ~65%. SEM study also confirms the morphology improvement with voids reduction, smoother/flatter surface, and increased grain size (see Supplementary Fig. 4 for SEM images of MAPbBr$_3$ thin films before and after NIL). Next, we quantify the surface roughness reduction by atomic force microscopy (AFM), as depicted in Figure

3a-b (bottom images). Before NIL, the film has a root-mean-square (RMS) roughness of 3.84nm from a 2μm×2μm area, which is smoother than typically reported values of around 30nm [36–38]. After NIL, the RMS roughness reduces by five-fold to 0.75nm.

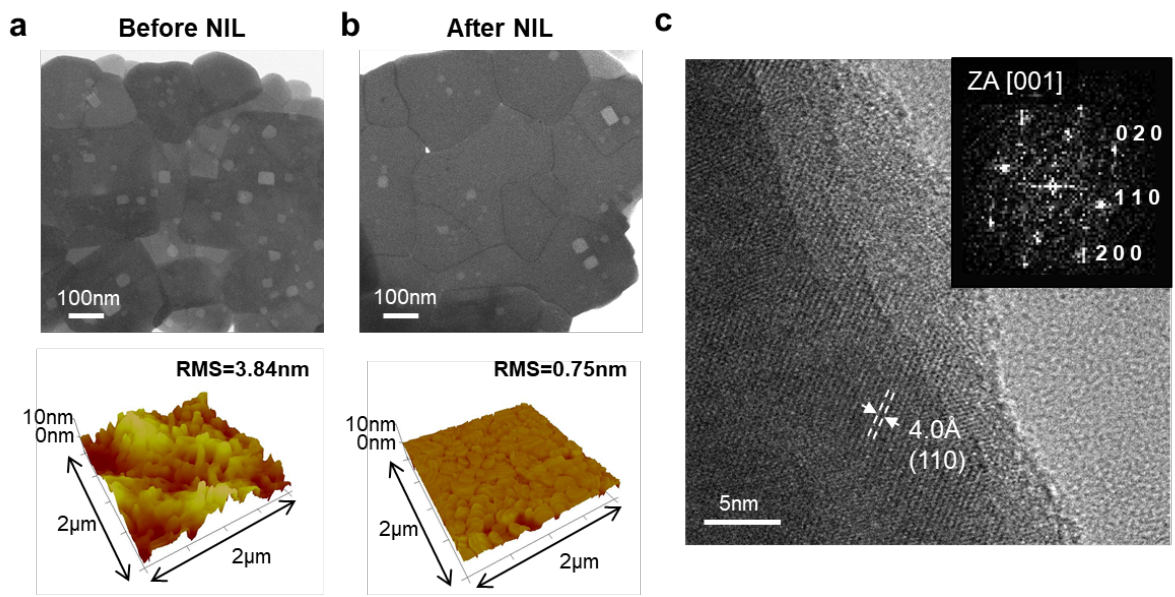

Figure 3 Direct observation of crystallinity and morphology enhancement. **a.** STEM (top) and AFM (bottom) of MAPbBr$_3$ thin film before NIL. **b.** STEM (top) and AFM (bottom) of MAPbBr$_3$ thin film after NIL. **c.** TEM of MAPbBr$_3$ thin film after NIL (inset: FFT pattern with [001] zone axis).

To perform crystallographic analysis on MAPbBr$_3$, we carry out high resolution TEM (HRTEM) measurements. The HRTEM image shown in Figure 3c depicts the arrangement of atoms in the nanoimprinted MAPbBr$_3$, showing an interplanar spacing of 4Å, indexable to the (110) planes of a cubic structure. The inset of Figure 3c shows the fast fourier transform (FFT) pattern with the [001] zone axis of MAPbBr$_3$. To assess the crystal structure and crystallinity of MAPbBr$_3$, we perform out-of-plane X-ray diffraction (XRD) measurements. Figure 4a shows the XRD spectra of MAPbBr$_3$ before and after NIL. Only three sharp diffraction peaks can be seen: they are at 2θ=14.93°, 30.13°, 45.89°, corresponding to the (001), (002), and (003) planes, respectively. Although the XRD result suggests the material to be single cubic crystal, morphology examinations by STEM, AFM (Figure 3a-b) and SEM (Supplementary Fig. 4a-b) prove its poly-crystal nature. These studies allow us to conclude that our MAPbBr$_3$ has highly oriented crystals in the surface-normal direction and randomly distributed crystals in the lateral surface direction. After NIL, the film has ~2 times higher count in the (00$l$) plane, implying crystallinity improvement in the surface-normal [001] direction.

To distinguish between the single crystal and highly oriented poly-crystal quality of our material, we perform pole figure XRD measurements of reflections from the (002) planes. By monitoring the diffracted intensity with respect to the various crystal orientations in the specimen, we map all of the (002) plane crystallites' orientation in the 3-dimensional space. While the diffraction from a completely random orientation of a poly-crystalline material would result in the same intensity in all directions, the pole figure of our MAPbBr$_3$ (Figure 4b) shows that the surface-normal [001] direction has the highest intensity, forming a continuous ring shape in the lateral surface [100] direction. This indicates that the film has poly-crystallites with strongly preferred orientation in the surface-normal direction. However, the film has no preferred orientation in the lateral surface direction.

To quantify the degree of texture, we carry out high-resolution XRD (HRXRD) omega rocking curve measurements (Figure 4c). By rocking the angles through a range of omega values with a fixed 2θ=14.93° and detecting the (001) plane reflections, we measure the deviation angle from their average crystallographic orientations. Although a single crystal should have a 0° theoretical value; a value of 0.01877° is measured for single crystal Si (004), representing the instrument resolution. On the other hand, a polycrystalline material that has no texture or a low degree of texture would have an undefined angle or a large angle, respectively. With our MAPbBr$_3$ films, we measure a 1.9° deviation from the average crystallographic orientation before NIL, and a narrower 1.1° with higher intensity after NIL, thus supporting our conclusion of the improved crystallinity. Furthermore, these XRD results imply enhanced preferential growth of crystallization. These combined characterization results (STEM, AFM, TEM in Figure 3, SEM in Supplementary Fig. 4, and XRD in Figure 4) allow us to speculate the mechanism behind the crystallinity improvement by NIL. Figure 4d illustrates our speculated mechanism: the elevated pressure and temperature during NIL (see Supplementary Part A for a detailed description of our NIL process) stimulates the reorientation of crystals by restricting their growth in the surface-normal direction, resulting in a smoother/flatter surface, and higher crystallinity with larger grain size. As a result, we observe PL enhancement from MAPbBr$_3$ after flat NIL because of the improved film quality (see Supplementary Fig. 4c for PL data).

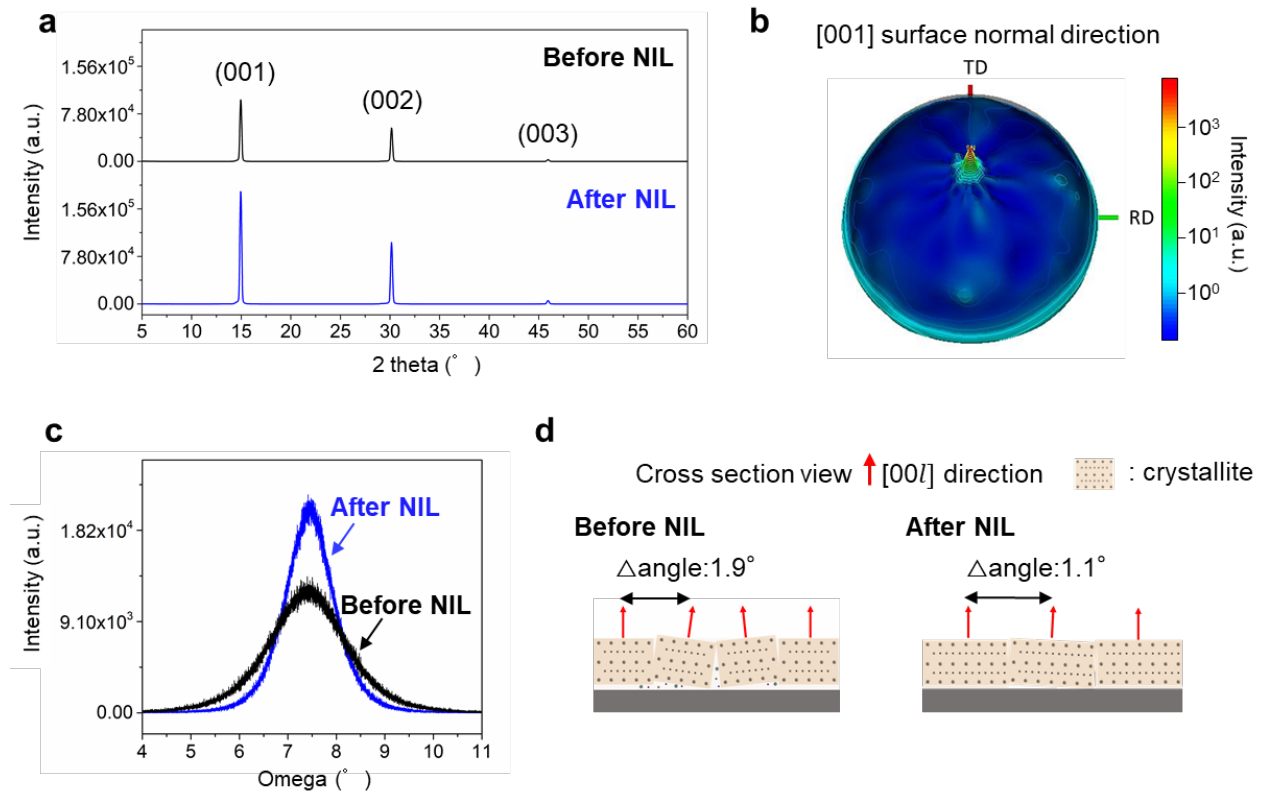

Figure 4 Characterization of bulk crystallinity and orientation before and after NIL. **a.** Out of plane measurement. **b.** Pole figure measurement in the surface-normal direction. TD: transverse direction, RD: rolling direction. **c.** Omega rocking curve measurement. **d.** Schematic of the NIL's effect.

Although NIL enhances the film quality thus the optical property, the imprinted MAPbBr$_3$ still contains defects and grain boundaries that serve as trigger points of degradation. We further improve the device quality by not only encapsulating MAPbBr$_3$, but also defect-passivating the trigger points with a thin layer of PC film. Compared to the popular choice of the ALD-Al$_2$O$_3$ encapsulation layer [39–42], we find PC encapsulation to be more suitable, thanks to its better uniformity, lower processing temperature, cost-effectivity, and fabrication simplicity (see Supplementary Part D for the study of encapsulation with ALD-Al$_2$O$_3$). In terms of defect passivation, more defects are present when ALD-Al$_2$O$_3$ is used in place of PC (see Supplementary Part D for SEM images).

To investigate the effect of PC defect passivation during laser operation, we treat the PC-passivated MAPbBr$_3$ thin films with more extreme conditions than our optical pumping condition. We illuminate the film with high photon energies of 4.8eV and 6.7eV from a low-pressure mercury lamp without any substrate cooling (see Supplementary Part E for details of this study). Figure 5 shows that without PC's protection, UV destroys perovskite grains. Compared to as-spin-coated sample in Figure 5a-top, more pinholes and a roughened surface are evident after the UV treatment (Figure 5a-bottom). On the other hand, once passivated with PC, perovskite grains are protected from UV (Figure 5b-top in comparison with Figure 5b-bottom). For nanoimprinted samples, PC plays a similar role (Figure 5c-top and bottom). We further analyze PC's effect by XRD: a specimen passivated by PC exhibits the same crystallinity before and after UV (Supplementary Fig. 7). Figure 5d depicts a simplified schematic of the degradation mechanisms by heat, photon energy, oxygen, and moisture (top), and grain protection by PC passivation of pinholes and grain boundaries (bottom).

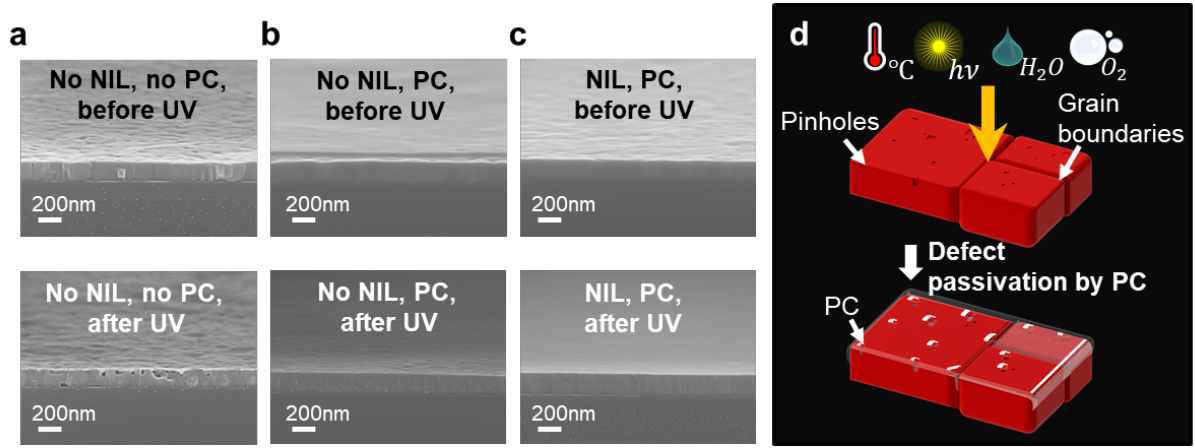

Figure 5 PC's defect passivation effect on MAPbBr$_3$. **a.** As-spin-coated film before (top) and after (bottom) UV. **b.** Spin-coated film with PC before (top) and after (bottom) UV. **c.** Nanoimprinted film with PC before (top) and after (bottom) UV. **d.** Schematic of PC passivation to prevent perovskite degradation from heat, photon energy from UV, moisture, and oxygen.

Last but not least, another advantage of the PC passivation layer is that due to the light curing effect, it also enhances perovskite's emissive properties (see Supplementary Part E for PL data). While we do observe the known light curing effect from the direct UV exposure on as-spin-coated perovskite [43,44], the emission enhancement is accompanied by a wavelength shift (Supplementary Fig. 8a). Furthermore, prolonged UV exposure destroys grains and in fact leads to the quenching of perovskite's emission (Supplementary Fig. 8b). In contrast, when PC-passivated perovskite is exposed to direct UV, perovskite grains do not get damaged; the emission wavelength remains the same; and the PL enhancement is more significant than the case without PC passivation (Supplementary Fig. 8c). It is known that UV polymerizes perovskite when the perovskite is embedded in polymer, i.e. photopolymerization takes places [45,46]. Therefore, an additional benefit of our defect passivation method is to promote the UV curing effect while maintaining perovskite's emission wavelength during optical pumping.

Lastly, we construct a perovskite diode structure and conduct electroluminescence (EL) measurements to investigate whether our strategies also improve device performance under electrical injection. To solely investigate our method's effect on the perovskite gain region without any influence from carrier transport layers, we construct single layer LEDs [47] which have a material stack of InGa (top electrode)/perovskite/ITO (bottom electrode) without hole injection layer (HIL) or electron injection layer (EIL). To investigate the effect of NIL alone, the un-structured $MAPbBr_3$ gain region is prepared by imprinting $MAPbBr_3$ thin film with a flat stamp. The resulting device has a high current efficiency of 0.076 cd/A at 3.5V, whereas the maximum current efficiency for the LED with as-spin-coated $MAPbBr_3$ gain is 0.018 cd/A at 4V (Supplementary Fig.10a-c). We also observe PC's effect on the current efficiency by applying a PC layer of various thicknesses (0.08%, 0.14%, and 0.24% concentrations, resulting in thicknesses of <5nm, ~7nm, ~10nm) on $MAPbBr_3$. Reduction of current density at any given voltage is expected when an insulating layer is inserted in a LED. However, the charge balance can be improved when applying its optimal thickness on top of perovskite [48]. As shown in Supplementary Fig.10d-f, the ~7nm thick PC-passivated device shows higher luminescence at 3.5V with higher current efficiency of 0.075 cd/A than devices prepared under other conditions. This study shows that optimal thickness PC on $MAPbBr_3$ leads to the similar current efficiency enhancement as NIL alone. When these two approaches are used together, a 7-fold enhancement of current efficiency is observed from PC passivated flat NIL $MAPbBr_3$ LED, reaching 0.126 cd/A (1584 cd/m$^2$) at 4V (Supplementary Fig.10g-i).

Conclusion

By directly forming high quality cavities and applying defect passivation, we have demonstrated stable CW lasing behavior at room temperature from $MAPbBr_3$ with a low threshold of 9.8W/cm$^2$ without any substrate cooling. We investigated the underlying mechanisms behind the superior device performance by comprehensively studying $MAPbBr_3$ thin film's morphology and crystallinity change during NIL and after PC defect passivation. We show that NIL can reduce the number of intergranular and intragranular defects, enlarge crystals, and lead to smoother/flatter surfaces. We further show that a thin layer of PC film can effectively passivate the remaining surface defects, and additionally, promote UV light curing without shifting the perovskite's emission wavelength. Through the combined analysis of the crystal quality, defect passivation, and laser performance, we provide crucial strategies to achieve stable CW room temperature operation of perovskite lasers. We also show that our approaches are effective for electrically driven structures such as LEDs, hence making the realization of electrically driven perovskite lasers within reach.

Methods

**Cavity fabrication with nanoimprint lithography (NIL).** $MAPbBr_3$ thin films were prepared on $SiO_2$(1um)/Si substrates by spin-coating solution ($MABr:PbBr_2$ in GBL:DMSO) and annealing at 70°C for 5mins. Next, NIL was performed with a pre-fabricated nanopillar stamp placed on top of the thin film. Subsequently, 0.52% PC solution was spun-coat on top of the patterned perovskite and annealed for 30mins at 60°C, to form a 50nm thick uniform layer of PC thin film. A more detailed description can be found in Supplementary Part A.

**Scanning electron microscopy (SEM).** Secondary electron SEM images were taken using an in-lens detector at an accelerating voltage of 10kV by a Zeiss Supra-40 SEM or a Raith150TWO e-beam lithography system. For insulating samples such as PC or $Al_2O_3$ coated samples, ~4nm of gold was sputtered (Hummer VI at 20mA, 120mTorr) before SEM examination.

**Atomic resolution scanning transmission electron microscopy (STEM)/ TEM.** For atomic resolution imaging, aberration corrected JEM-ARM200F (JEOL USA Inc.) was employed at an accelerating voltage of 200kV, which can be operated either STEM mode or TEM mode. The morphology of the thin films before and after NIL was investigated using the STEM mode and the atomic structure of nanoimprinted films was examined using the TEM mode.

To reduce the sample thickness for STEM/TEM examination, $MAPbBr_3$ thin films was scrapped by a razor blade, and the scrapped film was dispersed in toluene by ultrasonication for a few mins. The resulting solution (toluene and scrapped $MAPbBr_3$) was dropped on a TEM grid (Cu-300HD, Copper grids, 300 mesh), and STEM/TEM was performed after the toluene had evaporated. For nanoimprinted samples, $MAPbBr_3$ films were imprinted with a flat stamp before being scrapped with a razor blade. Note that although specimen preparation for TEM usually requires Focused Ion beam (FIB) to thin down the sample, the above approach was used to avoid damage during sample preparation [49].

**X-ray diffraction (XRD).** XRD measurements were performed using a Rigaku SmartLab X-ray Cu target (Ka1=1.54059 Å) and a HyPix 3000 detector. To see the thin film crystallinity and structure, out-of plane measurement (2-theta/omega scan) was performed on all samples under the same measurement condition, in the 2-theta range of 5° to 60° with 0.01° step, and ~2°/min scan speed.

The thin film texture was examined with the in-plane pole figure measurement configuration [50]. 2-theta in the 0°–360° range was scanned by rotating the sample (beta axis) with fixed 2-theta and alpha 3° steps (by tilting the detector around the sample instead of tilting the sample).

High resolution XRD (HRXRD) rocking curve measurements were performed with a high resolution monochromator (Ge(220) 2X). All samples were measured in the omega range of 3.4789°–11.4789° by rocking with a fixed 2-theta, with 0.001°step and 1.3°/min scan speed.

**X-ray photoelectron spectroscopy (XPS).** XPS was used for the elemental compositional analysis of the sample surface. XPS measurements were performed on perovskite thin films using a Versa Probe II at an ultrahigh vacuum of $10^{-8}$Pa. The X-ray source used is an Al K$\alpha$ which has 1486.6eV photon energy, 50W gun power, 15kV operating voltage, 200um X-ray spot size, and 59° angle between the X-ray source and the detector. Before obtaining the XPS data, calibration using an internal standard Au, Cu, and Ag samples was performed. XPS data was obtained with 6 sweeps at a pass energy of 187.85eV, and high resolution XPS data for each element was obtained with 6 sweeps at a pass energy of 23.5eV. Data was obtained without injecting a flux of low energy electrons. The peaks were monitored by overlapping the first and last sweep data to confirm that the data does not contain any surface charging – an issue that is usually associated with insulating samples. To examine the element distribution from MAPbBr$_3$ thin film surface as a function of distance, XPS depth profiling was performed with 3 sweeps at a pass energy of 23.5eV prior to each cycling of sputtering, using monatomic Ar at 1kV for 5mins.

**Atomic force microscopy (AFM).** To examine the morphology of thin films, AFM images were taken using a Veeco Multimode V SPM. The thin films were scanned for 2μmX2 μm area at 0.8Hz rate using an AFM tip (OTESPA-R3 from Bruker).

**Steady state micro-photoluminescence (micro-PL) spectroscopy.** All CW PL measurements were carried out at room temperature with 50% humidity. To locate the device (100μmX100μm in size) under test, a cascaded 4-f system is employed. Using the CW operation mode of a 355nm pump laser (TALON-355-20, Spectra-Physics), the samples were excited with a round beam of 20μm diameter via a 5X objective lens. The emission from the sample was collected using a cascaded 4-f system and focused onto the slit of a spectrograph (Princeton Instruments, IsoPlane SCT-320) coupled to a cooled Si detector (Princeton Instruments, PIXIS:400BRX). Alternatively, the emission was directed to a CCD camera (Ophir BM-USB-SP620) to capture the far field emission pattern. A more detailed description can be found in Supplementary Part F.

**EL measurement.** All measurements were conducted using a mechanical prob-station under high vacuum <20mTorr. Current density-Voltage (J-V) and Luminance-Voltage (L-V) characteristics were measured using a Keithley 2400 source meter and a Photo Research PR-650 spectroradiometer in the range of 0–5V with 0.5V increment. A more detailed description can be found in Supplementary Part G.

**Acknowledgements:** This work is supported by the UT Dallas faculty start-up fund, Welch Foundation (grant AT-1992-20190330 and AT-1617), National Science Foundation (No. CBET-1606141), and The University of Texas at Dallas Office of Research through the SPIRe Grant Program.